\documentclass{article}
\usepackage[utf8]{inputenc}
\usepackage[british]{babel}
\usepackage{hyperref}
\usepackage{underscore}
\usepackage{listings}
\usepackage{url}
\usepackage{amsthm}
\usepackage{tikz}
\usepackage[font=small,labelfont=bf]{caption}
\usepackage{algorithm}
\usepackage{algorithmicx}
\usepackage[noend]{algpseudocode}
\usepackage{array}

\usetikzlibrary{arrows}
\usetikzlibrary{positioning}

\theoremstyle{definition}
\newtheorem{definition}{Definition}[section]

\algdef{SE}[VARIABLES]{Variables}{EndVariables}
{\algorithmicvariables}
{\algorithmicend\ \algorithmicvariables}
\algnewcommand{\algorithmicvariables}{\textbf{Variables}}

\title{Functional Design of Computation Graph}

\author{Pierre Vandenhove}

\newcommand{\op}{\lambda}

\begin{document}
\lstset{language=Caml}
\tikzset{
	vertex/.style={circle,draw,minimum size=1em},
	edge/.style={->,> = latex'}
}
\maketitle

\abstract{
Representing the control flow of a computer program as a computation graph can bring many benefits in a broad variety of domains where performance is critical. This technique is a core component of most major numerical libraries (TensorFlow, PyTorch, Theano, MXNet,...) and is successfully used to speed up and optimise many computationally-intensive tasks. However, different design choices in each of these libraries lead to noticeable differences in efficiency and in the way an end user writes efficient code.

In this report, we detail the implementation and features of the computation graph support in OCaml's numerical library Owl, a recent entry in the world of scientific computing.
}

\section{Introduction} \label{sec:intro}
With the recent progress of deep learning methods in fields such as computer vision and natural language processing, the need for performant and user-friendly frameworks for scientific computing is especially high. In this context, \textit{computation graph} (sometimes referred to as \textit{dataflow graph}) support is an essential tool to achieve state-of-the-art performance. This report explains the way this feature is implemented in \textit{Owl}\footnote{\url{https://github.com/owlbarn/owl}}~\cite{owl}, an emerging numerical library developed with the OCaml language.

The main idea behind a computation graph is to replace eager evaluation of variables, which is the default behaviour in OCaml, by the building of a graph representing the dependencies between variables (scalar values or arrays) and the operations linking them. The actual evaluation of the outputs of the graph is delayed so that the graph can be manipulated to reduce memory footprint and improve performance before calling any computationally-heavy operation.

There are numerous benefits from using this approach:
\begin{itemize}
	\item preliminary optimisation of the graph structure to reduce the complexity of the computation and improve performance (see Section~\ref{sec:core});
	\item reduction of the memory management overhead by preallocating space for each node (see Section~\ref{sec:memo});
	\item reduction of the memory consumption by sharing allocated memory blocks between multiple nodes (see Section~\ref{sec:memo});
	\item natural support for incremental computation (recomputation of exactly what is needed when some input is modified);
	\item natural support for parallel and distributed computing (see Section~\ref{sec:devdep});
	\item natural support for symbolic mathematics.
\end{itemize}
This is especially relevant for a program that manipulates large multi-dimensional arrays (often called \textit{tensors}) and for iterative optimisation algorithms, because the cost of building, storing and optimising the graph becomes negligible.
These features also make it easier to write performant programs: a user simply has to write a high-level description of the control flow of the program. Everything else is automatically handled and optimised by Owl. This allows for quick and efficient prototyping.

\section{Definitions}

The following definition is inspired by \cite{le2018tflms}.
\begin{definition}[Computation graph] \label{def:cg}
	Let $O$ be a set of operations such that $\texttt{Var} \in O$.
	
	A \textit{computation graph} (or \textit{dataflow graph}) over $O$ is a graph $G = (V, E, \op, U)$, where $V$ is the set of vertices of $G$, $E \subseteq V \times V$ is the set of directed edges, $\op : V \rightarrow O$ is a function mapping each vertex to an operation $o \in O$ and $U \subseteq V \times V$ is the set of \textit{update} edges. We require that the graph $(V, E)$ is acyclic and that for all $(u, v) \in U$, $\op(v) = \texttt{Var}$.
\end{definition}

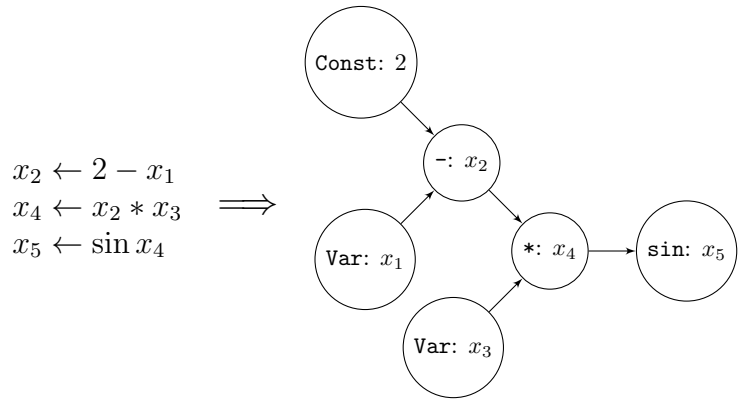
\begin{figure}[!ht]
\centering
\begin{tabular}{m{.23\textwidth}cm{0.5\textwidth}}
	\centering
	\begin{minipage}{0.23\textwidth}
	\large{\begin{algorithmic}
		\State $x_2 \gets 2 - x_1$
		\State $x_4 \gets x_2 * x_3$
		\State $x_5 \gets \sin x_4$
	\end{algorithmic}}
	\end{minipage} &
	\Large{$\Longrightarrow$} &
	\begin{tikzpicture}[scale=0.9, every node/.style={transform shape}]
	\node[vertex] (x5) at (0,0)           {\texttt{sin}: $x_5$};
	\node[vertex] (x4) [left = .7cm of x5] {\texttt{*}: $x_4$};
	\node[vertex] (x3) [below left = .7cm of x4] {\texttt{Var}: $x_3$};
	\node[vertex] (x2) [above left = .7cm of x4] {\texttt{-}: $x_2$};
	\node[vertex] (x1) [below left = .7cm of x2] {\texttt{Var}: $x_1$};
	\node[vertex] (c)  [above left = .7cm of x2] {\texttt{Const}: $2$};
	
	\foreach \from/\to in {c/x2,x1/x2,x2/x4,x3/x4,x4/x5}
	\draw[edge] (\from) -- (\to);
	\end{tikzpicture}
\end{tabular}
\caption{A simple program and its computation graph.\label{fig:cg1}}
\end{figure}

Each vertex of the computation graph represents an operation. An edge $(u, v) \in E$ means that the output of vertex $u$ will be used as an input by the operation of vertex $v$ --- it defines the dependencies between the operations. With a distributed system, different vertices of the graph might be computed on different machines. The graph is then also a description of the network linking the vertices together. You can see an example of a program and its computation graph in Figure~\ref{fig:cg1}. In this example, to evaluate the output vertex $x_5$, we need to specify values for the variables $x_1$ and $x_3$. If after evaluating it once, we only modify the value of $x_3$, there is no need to re-evaluate $x_2$ (this is an example of incremental computation).

The set of \textit{update edges $U$} is a mechanism to allow reusing the value of some vertices at the end of an evaluation as variables for the next evaluation. This is necessary to express recurrent neural networks or neural network training (where at the end of each iteration, weights are updated). Note that the graph $(V, E \cup U)$ can contain cycles. Notice also that the graph is not necessarily \textit{simple} in the graph-theoretical meaning: two vertices can be linked by more than one edge (for instance with the computation $x \gets y * y$).

\begin{definition}[Topological Ordering]
	Let $G = (V, E, \op, U)$ be a computation graph with $n$ vertices. A \textit{topological ordering} of $G$ is a bijection $\gamma : V \rightarrow \{0, 1, \ldots, n - 1\}$ such that for all $(u, v) \in E$, $\gamma(u) < \gamma(v)$.
\end{definition}

The vertices of a computation graph need to be evaluated in any topological order.

\section{Memory allocation problem}
Allocating memory to each vertex of a computation graph is a problem similar to one that was first described in 1973~\cite{pebbles} to look for an efficient algorithm for register allocation, using an abstraction called the \textit{pebble game}. We recall its definition.

\begin{definition}[Pebble game] \label{def:pebble}
	The \textit{pebble game} is played on a directed acyclic graph (DAG). Each vertex can store at most one pebble. The game begins with no pebble on any vertex. At each step, the player can perform one of the following moves:
	\begin{enumerate}
		\item if a vertex $v$ has no predecessor (\textit{input vertex}), the player can place a pebble on $v$.
		\item if all predecessors of a vertex $v$ are pebbled, the player can place a pebble on $v$ or \textit{slide} a pebble from one of its predecessors to $v$.
		\item the player can remove any pebble from a vertex (and reuse that pebble later).
	\end{enumerate}
	The goal of the game is to place a pebble at least once on some fixed \textit{output vertices} of the graph. A \textit{pebbling strategy} is a sequence of moves following the rules of the game and reaching the goal. The \textit{space} used by a pebbling strategy is the maximum number of pebbles used during the execution of the strategy. The \textit{time} of a strategy is the number of times a pebble is placed on a vertex (without counting the removals of pebbles).
\end{definition}

This relates to the memory allocation of the computation graph if we see pebbles as memory blocks used to store the output value of a vertex. We assume that the values of the \textit{input vertices} are known (move 1). We can only compute the value of a vertex if all its predecessors are simultaneously stored in memory (move 2). The \textit{sliding} move means that the memory of a vertex can be overwritten by its successor during its computation (\textit{inplace reuse}). We can always reuse a memory block from another vertex (move 3).

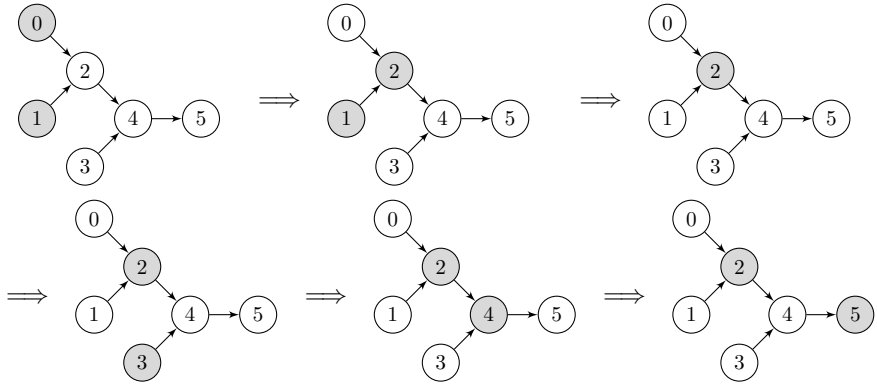
\begin{figure}[!ht]
	\begin{tabular}{m{3cm}cm{3cm}cm{3cm}}
		\centering
		\begin{tikzpicture}[scale=0.8, every node/.style={transform shape}]
		\node[vertex] (x5) at (0,0)           {$5$};
		\node[vertex] (x4) [left = .5cm of x5] {$4$};
		\node[vertex] (x3) [below left = .5cm of x4] {$3$};
		\node[vertex] (x2) [above left = .5cm of x4] {$2$};
		\node[vertex,fill=gray!30] (x1) [below left = .5cm of x2] {$1$};
		\node[vertex,fill=gray!30] (c)  [above left = .5cm of x2] {$0$};
		\foreach \from/\to in {c/x2,x1/x2,x2/x4,x3/x4,x4/x5}
		\draw[edge] (\from) -- (\to);
		\end{tikzpicture}
		& $\Longrightarrow$ &
		\begin{tikzpicture}[scale=0.8, every node/.style={transform shape}]
		\node[vertex] (x5) at (0,0)           {$5$};
		\node[vertex] (x4) [left = .5cm of x5] {$4$};
		\node[vertex] (x3) [below left = .5cm of x4] {$3$};
		\node[vertex,fill=gray!30] (x2) [above left = .5cm of x4] {$2$};
		\node[vertex,fill=gray!30] (x1) [below left = .5cm of x2] {$1$};
		\node[vertex] (c)  [above left = .5cm of x2] {$0$};
		\foreach \from/\to in {c/x2,x1/x2,x2/x4,x3/x4,x4/x5}
		\draw[edge] (\from) -- (\to);
		\end{tikzpicture}
		& $\Longrightarrow$ &
		\begin{tikzpicture}[scale=0.8, every node/.style={transform shape}]
		\node[vertex] (x5) at (0,0)           {$5$};
		\node[vertex] (x4) [left = .5cm of x5] {$4$};
		\node[vertex] (x3) [below left = .5cm of x4] {$3$};
		\node[vertex,fill=gray!30] (x2) [above left = .5cm of x4] {$2$};
		\node[vertex] (x1) [below left = .5cm of x2] {$1$};
		\node[vertex] (c)  [above left = .5cm of x2] {$0$};
		\foreach \from/\to in {c/x2,x1/x2,x2/x4,x3/x4,x4/x5}
		\draw[edge] (\from) -- (\to);
		\end{tikzpicture}
	\end{tabular}
	
	\begin{tabular}{cm{2.7cm}cm{2.7cm}cm{2.7cm}}
		\centering
		$\Longrightarrow$ &
		\begin{tikzpicture}[scale=0.8, every node/.style={transform shape}]
		\node[vertex] (x5) at (0,0)           {$5$};
		\node[vertex] (x4) [left = .5cm of x5] {$4$};
		\node[vertex,fill=gray!30] (x3) [below left = .5cm of x4] {$3$};
		\node[vertex,fill=gray!30] (x2) [above left = .5cm of x4] {$2$};
		\node[vertex] (x1) [below left = .5cm of x2] {$1$};
		\node[vertex] (c)  [above left = .5cm of x2] {$0$};
		\foreach \from/\to in {c/x2,x1/x2,x2/x4,x3/x4,x4/x5}
		\draw[edge] (\from) -- (\to);
		\end{tikzpicture}
		& $\Longrightarrow $ &
		\begin{tikzpicture}[scale=0.8, every node/.style={transform shape}]
		\node[vertex] (x5) at (0,0)           {$5$};
		\node[vertex,fill=gray!30] (x4) [left = .5cm of x5] {$4$};
		\node[vertex] (x3) [below left = .5cm of x4] {$3$};
		\node[vertex,fill=gray!30] (x2) [above left = .5cm of x4] {$2$};
		\node[vertex] (x1) [below left = .5cm of x2] {$1$};
		\node[vertex] (c)  [above left = .5cm of x2] {$0$};
		\foreach \from/\to in {c/x2,x1/x2,x2/x4,x3/x4,x4/x5}
		\draw[edge] (\from) -- (\to);
		\end{tikzpicture}
		& $\Longrightarrow$ &
		\begin{tikzpicture}[scale=0.8, every node/.style={transform shape}]
		\node[vertex,fill=gray!30] (x5) at (0,0) {$5$};
		\node[vertex] (x4) [left = .5cm of x5] {$4$};
		\node[vertex] (x3) [below left = .5cm of x4] {$3$};
		\node[vertex,fill=gray!30] (x2) [above left = .5cm of x4] {$2$};
		\node[vertex] (x1) [below left = .5cm of x2] {$1$};
		\node[vertex] (c)  [above left = .5cm of x2] {$0$};
		\foreach \from/\to in {c/x2,x1/x2,x2/x4,x3/x4,x4/x5}
		\draw[edge] (\from) -- (\to);
		\end{tikzpicture}
	\end{tabular}
	\caption{Optimal pebbling of a computation graph.\label{easypebbling}}
\end{figure}

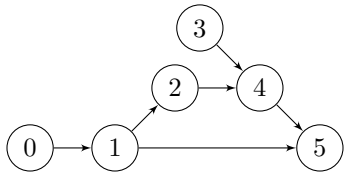
\begin{figure}[ht!]
	\centering
	\begin{tikzpicture}
	\node[vertex] (5) at (0,0) {5};
	\node[vertex] (4) [above left = .5cm of 5] {4};
	\node[vertex] (3) [above left = .5cm of 4] {3};
	\node[vertex] (2) [left = .5cm of 4] {2};
	\node[vertex] (1) [below left = .5cm of 2] {1};
	\node[vertex] (0) [left = .5cm of 1] {0};
	\foreach \from/\to in {0/1,1/5,1/2,2/4,3/4,4/5}
	\draw[edge] (\from) -- (\to);
	\end{tikzpicture}
	\caption{One pebbling strategy cannot reach the minimal values of both space and time. \label{hardpebbling}}
\end{figure}

By pebbling a graph in topological order and removing pebbles when they are not needed anymore, it is always possible to pebble each vertex exactly once (time is minimal). However, such a strategy may not always yield a minimal space value.
If we consider the example from Figure~\ref{fig:cg1}, we notice that we can pebble it with a space of 2 and a time of 6 (see Figure~\ref{easypebbling}).
The values of space and time of this strategy are both minimal. We now consider another example in Figure~\ref{hardpebbling}. It is possible to pebble this graph with a time of 6 and a space of 3 (by leaving a pebble on vertex 1 until the end). It is also possible to pebble it with a time of 8 and a space of 2 (by repebbling 0 and 1 once 4 is pebbled). However, one single strategy cannot be minimal for both time and space.

We see that there is a time-space tradeoff and that depending on the space constraints, it might be worth computing the same vertex more than once. If a graph has certain properties, some algorithms can obtain interesting bounds on space complexity without sacrificing much of the time complexity. For instance, \cite{sqrtbound} uses the specific shape of the computation graph when training a neural network to devise a memory allocation strategy with a square root bound on space complexity by means of two computations of the forward pass. You can find more examples in \cite[Chapter~10]{modelscomputation}: there even exists a family of graph $G_k$ such that using $k$ pebbles takes exponential time at best, but using $k + 1$ pebbles takes minimal time. This illustrates that excessive memory minimisation is not always desired in practice.

The problem of whether there exists a pebbling strategy using less than $k$ pebbles has been proven to be \textsc{PSPACE}-complete~\cite{pspacepebbles}. If we do not allow any repebbling of the same vertex, the same problem is \textsc{NP}-complete~\cite{pebbles}, and even hard to approximate within any constant factor~\cite{oneshotpebble}. Since computation graphs can have a few thousand vertices, we will not consider exact algorithms to allocate memory.

For more information on pebbling games, see \cite{modelscomputation, nor}.

\section{Core implementation} \label{sec:core}
\subsection{Design}
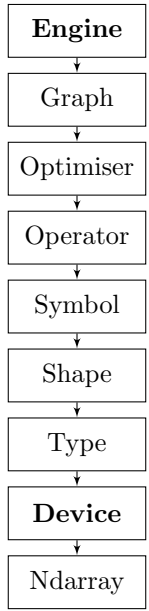
\begin{figure}[ht]
	\centering
	\begin{tikzpicture}
	[myrectangle/.style={rectangle, draw, minimum width=52, minimum height=20}]
	
	\node[myrectangle] (a) at (0,0) {\textbf{Engine}};
	\node[myrectangle, below=0.2 of a] (b) {Graph};
	\node[myrectangle, below=0.2 of b] (c) {Optimiser};
	\node[myrectangle, below=0.2 of c] (d) {Operator};
	\node[myrectangle, below=0.2 of d] (e) {Symbol};
	\node[myrectangle, below=0.2 of e] (f) {Shape};
	\node[myrectangle, below=0.2 of f] (g) {Type};
	\node[myrectangle, below=0.2 of g] (h) {\textbf{Device}};
	\node[myrectangle, below=0.2 of h] (i) {Ndarray};
	\foreach \from/\to in {a/b,b/c,c/d,d/e,e/f,f/g,g/h,h/i}
	\draw[edge] (\from) -- (\to);
	\end{tikzpicture}
	\caption{Owl's computation graph functor stack. \label{fig:stack}}
\end{figure}

Owl's computation graph\footnote{The code for Owl's computation graph implementation is available at \url{https://github.com/owlbarn/owl/tree/master/src/base/compute}.} has been implemented with the following ideas in mind:
\begin{itemize}
	\item it should be easy to switch between computation graph and standard eager evaluation --- both should have similar interfaces;
	\item it should be flexible enough to be easily extended to other devices and number types.
\end{itemize}
This indicates that there should be a core code used by all devices to build and represent generic computation graphs and that it should be separated as much as possible from device-dependant code --- each supported device should simply reimplement what is necessary to initialise data and perform the operations.
This is done by wrapping the tensor module (\texttt{Ndarray}) directly into a new functor stack, which you can see in Figure~\ref{fig:stack}.

Owl's computation graph is parameterised by a \textit{device} and a \textit{number type}. As an example, to use the computation graph on a CPU device with single-precision floating-point data type, you should replace
\begin{lstlisting}
module N = Dense.Ndarray.S
\end{lstlisting}
by
\begin{lstlisting}
module N =
  Owl_computation_cpu_engine.Make(Dense.Ndarray.S)
\end{lstlisting}
This module declaration is a abbreviation of
\begin{lstlisting}
module N =
  Owl_computation_engine.Flatten (
    Owl_computation_cpu_engine.Make_Nested (
      Owl_computation_graph.Make (
        Owl_computation_optimiser.Make (
          Owl_computation_operator.Make (
            Owl_computation_symbol.Make (
              Owl_computation_shape.Make (
                Owl_computation_type.Make (
                  Owl_computation_cpu_device.Make (
                    Dense.Ndarray.S)))))))))
\end{lstlisting}
which is the definition of the functor stack of Figure~\ref{fig:stack}.
This module \texttt{N} can be used as an functor argument in various parts of Owl such as the algorithmic differentiation module (\texttt{Owl_algodiff_generic.Make(N)}), which is itself used by the neural network module. The \texttt{Device} and \texttt{Engine} layers are device-dependant (see Section \ref{sec:devdep}). We describe the role of each core layer in Section~\ref{sec:stack}.

This implementation of the computation graph is \textit{static} in the sense that once a graph is built, it is not meant to be changed. This prevents the use of control structures such as \texttt{if ...\ else ...} that could lead to a different graph depending on some condition. The advantage is that a static approach yields a performance improvement over a dynamic one: the graph does not have to be rebuilt at every evaluation. This design choice relies on the assumption that the exact same graph has to be evaluated many times with different inputs, which turns out to be the case in many practical applications.

\subsection{Functor stack}\label{sec:stack}
\paragraph{Type}
The \texttt{Type} layer defines the different supported operations as a variant type. An operation can be parametrised by arguments (for instance, \texttt{Conv2d of padding * int array}). If an operation is not supported, user-defined functions on tensors can be used by defining them using eager evaluation and wrapping them in a \texttt{Delay} operation. As a rule of thumb, it is better to split a function in as many nodes as possible to allow for better memory optimisation and decrease temporal memory.

\paragraph{Shape}
This layer defines functions to infer the shape of the output value of each operation, given the shape of its inputs. Users only have to specify the shape of the inputs (usually \texttt{Var} and \texttt{Const} operations) and that information is propagated to infer the shapes of all operations. This is necessary to preallocate the memory of each node.

\paragraph{Symbol}
The \texttt{Symbol} layer provides functions to manipulate symbols as nodes of the computation graph.

\paragraph{Operator}
The \texttt{Operator} functor defines the functions to create nodes and their dependencies for all the operators defined in \texttt{Type}. These are the functions that should be called when writing numerical code using a computation graph. The names and arguments of the functions are voluntarily the same as their counterpart in the \texttt{Ndarray} module to allow for an easy switching between both interfaces. It overloads the standard eager evaluation with graph-building (thus lazy) functions.

\paragraph{Optimiser} \label{sec:opti}
The \texttt{Optimiser} layer defines several structural patterns that can be optimised or removed before any computation is performed. This reduces the size of the computation graph and thus the time and memory to evaluate it. Here is a non-exhaustive list of the optimisations that it can perform.
\begin{itemize}
	\item Nodes whose value only depends on constant nodes can be evaluated once and for all when the graph is created, and the constant nodes can be removed (\textit{constant folding}, see Figure~\ref{constfold}).
	\begin{figure}[ht!]
		\centering
		\includegraphics[width=0.8\textwidth]{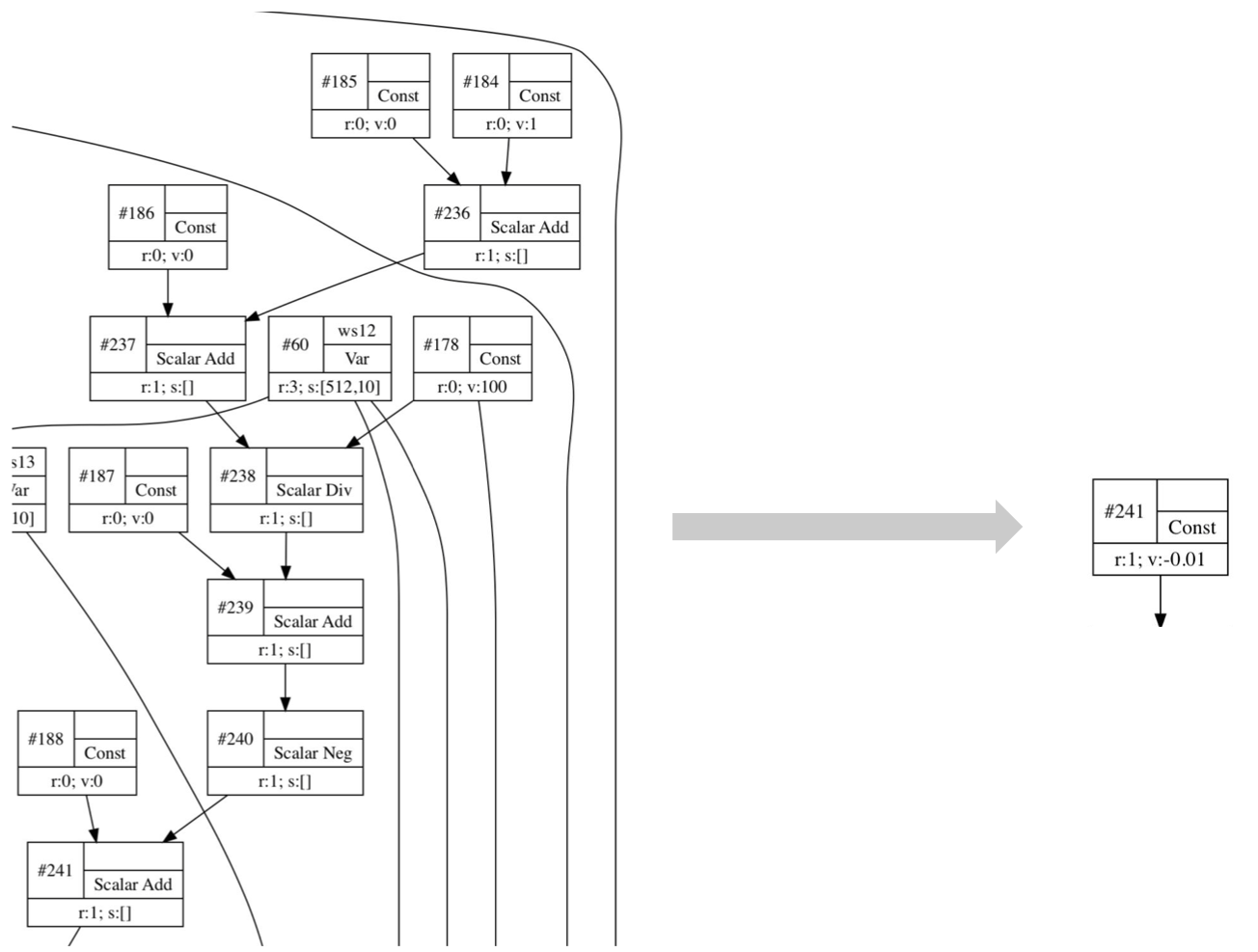}
		\caption{Constant folding: node \#241 depends only on constant nodes. \label{constfold}}
	\end{figure}
	\item Some patterns of nodes can be fused in order to be executed in a more efficient way. For instance, an \texttt{FMA} (\textit{fused-multiply-add}) node can replace a multiplication followed by an addition to reduce memory access and improve numerical accuracy. A common subpattern of \textsc{AdaGrad}~\cite{adagrad}, which is heavily used in neural network training, can also be fused into one node (see Figure~\ref{fig:fuse}).
	\begin{figure}[ht!]
		\centering
		\includegraphics[width=0.8\textwidth]{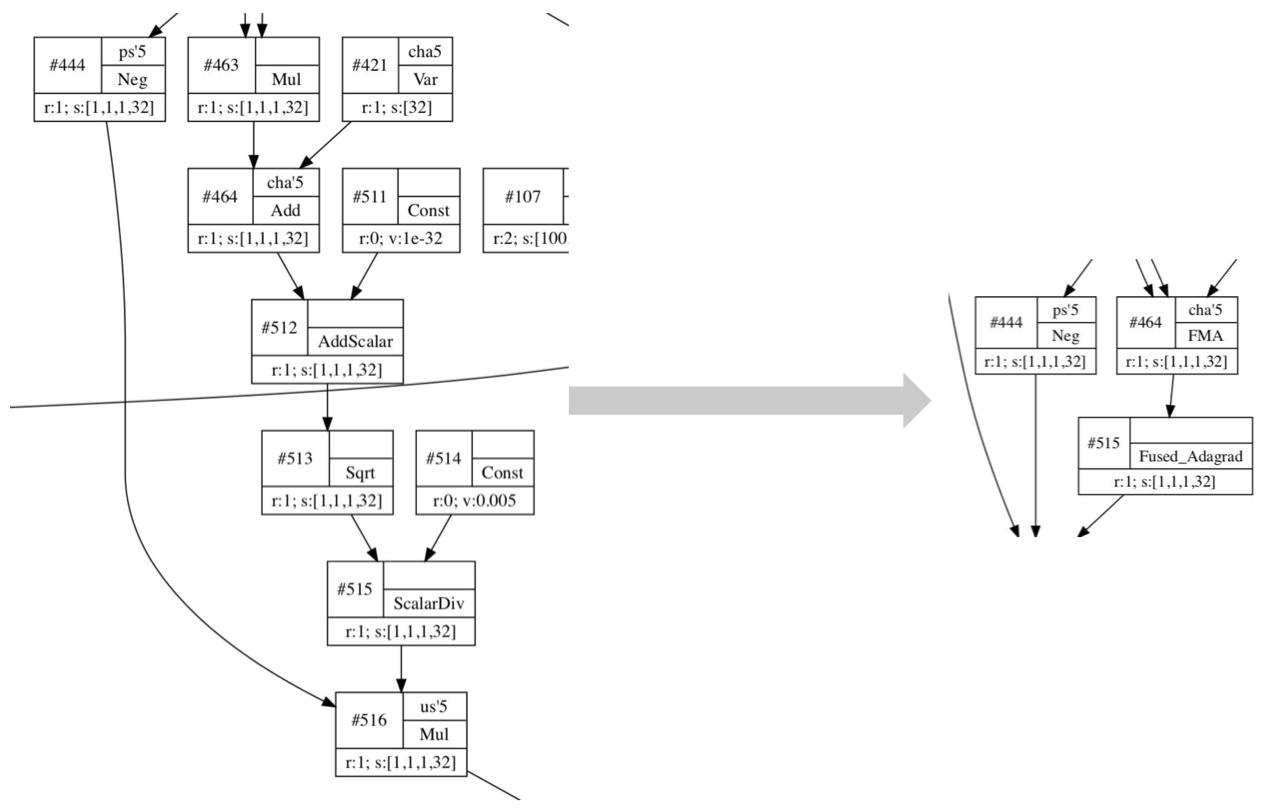}
		\caption{Merging of \#421, 463, 464 into \texttt{FMA} and of \#511-516 into \texttt{Fused_Adagrad}. \label{fig:fuse}}
	\end{figure}
	\item Useless calculations, such as adding zero, dividing by one, multiplying by zero or one, repeating the input of an operation supporting broadcasting,... are automatically optimised.
\end{itemize}

\paragraph{Graph}
This layer defines a \texttt{graph} type wrapping all the previous layers together. A \texttt{graph} is defined by its inputs and outputs nodes (from which the whole graph can be traversed). Inputs must be \texttt{Var} nodes and can be modified by the user before each evaluation. Graph visualisation functions are also provided. The mechanism to express the set of update edges $U$ from Definition~\ref{def:cg} is a variable called \texttt{iopair} which should be specified when the graph is built. The function \texttt{update_iopair} activating this mechanism should be called at the end of each graph evaluation.

\section{Device-dependant code} \label{sec:devdep}
Owl's computation graph currently supports both CPU and GPU devices. The GPU engine\footnote{\url{https://github.com/owlbarn/owl/tree/master/src/opencl}} is based on OpenCL~\cite{opencl}. We outline here the necessary parts to permit computation graph evaluation on a device.

\subsection{Device}
At the bottom of the functor stack, the \texttt{Device} layer abstracts implementation details about how tensors and scalar values are handled by the device. This is straightforward to implement for CPU devices --- a tensor is simply a contiguous block in memory. On GPUs\footnote{\url{https://github.com/owlbarn/owl/blob/master/src/opencl/compute/owl_computation_opencl_device.ml}}, tensors require a host (CPU) memory and a device (GPU) memory, plus a way to synchronise them.

\subsection{Initialisation} \label{sec:memo}
Each device stores values in a different way, thus memory initialisation is device-dependant. We have seen that the ability to preallocate memory for each vertex is one of the main advantages of using a computation graph. Compared to the pebble game (Definition~\ref{def:pebble}), we need to be careful of a few more caveats.
\begin{itemize}
	\item Some operations, dependently on how they are implemented, can not overwrite some of their parents while they are being carried out. In the pebble game, this corresponds to forbidding the sliding of one pebble for some operators.
	\item We need to take into account that some vertices have different output sizes. In the pebble game, this means that we want to assign to each pebble a size that must be larger than the output size of the vertices the pebble is placed on. What we actually want to minimise is the sum of the sizes of the pebbles rather than the number of pebbles (this is still equivalent to the original problem in the specific case where all the output sizes are equal).
	\item We want to always keep the value of some vertices in memory for practical purposes (for example, vertices $v$ such that $\op(v) = \texttt{Const}$, output vertices, vertices $u$ such that $(u, v) \in U$). In Owl, a boolean value is used to indicate whether the memory of a vertex can be reused.
\end{itemize}

We explain in more detail the memory allocation performed in Owl when running the code on a CPU device. For practical reasons, since the algorithm should be efficient for arbitrary graphs, we assume that each vertex can only be computed once (\textit{i.e.} keeping a minimal time value).
The chosen algorithm is inspired by the one used by MXNet~\cite{mxnet}.

A first challenge is to find an efficient way to share memory between vertices which do not have the same output size. Since Owl's multi-dimen\-sional array module is based on OCaml's \texttt{Bigarray}, we can use the \texttt{reshape} and \texttt{sub_left} functions in the following way:
\begin{lstlisting}[basicstyle=\footnotesize]
module N = Owl.Dense.Ndarray.S
let block = N.empty [|block_size|] in
let memory =
  N.reshape (N.sub_left block 0 node_numel) shape
\end{lstlisting}
The \texttt{block} variable is a one-dimensional array of sufficient size. We reuse the \texttt{node_numel} first elements of \texttt{block} and reshape that one-dimensional array to the right \texttt{shape}. Multiple nodes can thus base their memory on the same block, as long as their size is smaller than the size of the block, with no loss of performance.

The topological ordering $\gamma$ we use for allocation and evaluation is given by traversing the graph from the outputs using a post-order DFS.

By going through $\gamma$ in order, we can notice when the memory of a vertex becomes useless by keeping a counter of the number of times its value has been used by one of its successors. When all its successors have been evaluated, the memory block of the vertex can be tagged as reusable.

We follow the rules below to allocate a block of memory to a vertex $v$:
\begin{itemize}
	\item we only allocate a new block when no block is available;
	\item when multiple blocks are available, we pick the smallest block that is big enough to contain the value of $v$, so that as little memory as possible stays unused. If the available blocks are all smaller that $v$, we increase the size of the current biggest block to fit $v$;
	\item when possible, we always favour an \textit{inplace reuse} (\textit{i.e.} reusing the memory of a direct predecessor of $v$) to reduce memory access overhead.
\end{itemize}
You can find the pseudo-code of the algorithm in Algorithm~\ref{alg:allocate}.
\begin{algorithm} \caption{Memory Allocation for CPU devices\label{alg:allocate}}
	\begin{algorithmic}[noend]
		\Variables
		\State $refs$, associating to each node its number of successors
		\State $reusable$, used to store the available blocks
		\State $block\_size$, associating to each block a size
		\State $block$, associating to each node a block
		\EndVariables
		\\
		\Function{FindBestBlock}{$s$}
		\If{there is a block in $reusable$ with size $\geq s$}
		\State \Return smallest block in $reusable$ of size $\geq s$
		\ElsIf{$reusable$ is not empty}
		\State $b \gets$ largest block in $block$
		\State $block\_size(b) \gets s$
		\State \Return $b$
		\Else
		\State $b \gets$ new block
		\State $block\_size[b] \gets s$
		\State \Return $b$
		\EndIf 
		\EndFunction
		\\
		\Function{Initialise}{$x$}
		\begin{flushleft}
			\textbf{Input:} a node $x$ of the computation graph\\
			\textbf{Effect:} allocates a block of memory to $x$ and its ancestors.
		\end{flushleft}
		\If{$x$ is not initialised}
		\ForAll{predecessor $p$ of $x$} \textsc{Initialise}($p$)
		\EndFor
		\ForAll{predecessor $p$ of $x$}
			\State $ref[p] \gets ref[p] - 1$
			\If{$ref[p] = 0$} $reusable.add(p)$ \EndIf
		\EndFor
		\State $block[x] \gets$ \textsc{FindBestBlock($size(x)$)}
		\EndIf
		\EndFunction	
	\end{algorithmic}
\end{algorithm}

The overall time complexity of the allocation is $\mathcal{O}(n *  \log(b))$, where $n$ is the number of nodes in the graph and $b$ is the number of distinct memory blocks at the end of the algorithm (of course, $b \leq n$). The $\log(b)$ factor comes from the \textsc{FindBestBlock} function.

\subsection{Evaluation}
The device engine should also provide evaluation functions for all the operators defined in the \texttt{Type} functor, using functions defined in the selected \texttt{Ndarray} module. 

\section{Benchmarks}
We show in Figure~\ref{fig:stats} the gain that can be obtained by using a computation graph in a few notorious neural network architectures, namely InceptionV3~\cite{inception}, ResNet~\cite{resnet}, Mask R-CNN~\cite{mrcnn} (ported from implementation~\cite{mrcnnimp}) and a small neural network solving the MNIST number classification~\footnote{The implementation is available at \url{https://github.com/owlbarn/owl/blob/master/examples/lazy_mnist.ml}.}.

\begin{figure}[ht]
	\centering
	\footnotesize{
		\begin{tabular}{|l|c|c|c|c|c|}	
			\hline
			Architecture & Time w/o & \multicolumn{2}{c|}{Time w/ CG (s)} & Memory w/o & Memory w/ \\ \cline{3-4}
			& CG (s) & Building & Evaluating & CG (MB) & CG (MB) \\ \hline
			InceptionV3 & 0.5649 & 0.1066 & 0.2275 & 625.76 & 230.10 \\ \hline
			ResNet50 & 0.7933 & 0.1395 & 0.6090 & 1309.9 & 397.07 \\ \hline
			MNIST (training) & 20.422 & 0.1436 & 10.920 & 3685.3 & 895.32 \\ \hline
			Mask R-CNN & 11.538 & 0.3630 & 8.379 & 6483.4 & 870.48 \\ \hline
		\end{tabular}
	}
	\caption{InceptionV3 and ResNet50 are tested with a 299x299 image; Mask R-CNN is tested with a 768x768 image. The time is the average over 30 evaluations, without reusing pre-computed nodes when a computation graph is used. The graph building phase includes graph construction, optimisation and memory initialisation. The memory is the maximum resident set size of the program. This was evaluated on a laptop with an Intel i5-6300HQ CPU and 8 GB of RAM. \label{fig:stats}}
\end{figure}

\clearpage
\bibliography{cgbib} 
\bibliographystyle{alpha}
\end{document}